 \documentclass[structabstract]{aa}

\usepackage{graphicx}
 \usepackage[]{natbib}
 \usepackage{threeparttable}
\usepackage{txfonts}
\usepackage{url}
\bibpunct{(}{)}{;}{a}{}{,} 
\begin{document}

   \title{High resolution spectroscopic analysis of seven giants in the bulge globular cluster NGC 6723}

\author{A.~Rojas-Arriagada \inst{\ref{inst1}}
\and M.~Zoccali \inst{\ref{inst2}, \ref{inst3}}
\and S. V\'asquez \inst{\ref{inst2},\ref{inst3},\ref{inst4}}
\and V. Ripepi \inst{\ref{inst5}}
\and I. Musella \inst{\ref{inst5}}
\and M. Marconi \inst{\ref{inst5}}
\and A. Grado \inst{\ref{inst5}}
\and L. Limatola \inst{\ref{inst5}}
}
\institute{
Laboratoire Lagrange (UMR 7293), Universit\'e  Nice Sophia Antipolis, CNRS, Observatoire de la C\^ote d'Azur, CS 34229, 06304 Nice, cedex 04, France  \email{arojas@oca.eu} \label{inst1}
\and
Instituto de Astrof\'{i}sica, Facultad de F\'{i}sica, Pontificia Universidad Cat\'olica de Chile, Av. Vicu\~na Mackenna 4860, Santiago, Chile \label{inst2}
\and
The Milky Way Millennium Nucleus, Av. Vicu\~{n}a Mackenna 4860, 782-0436 Macul, Santiago, Chile\label{inst3}
\and
European Southern Observatory, A. de Cordova 3107, Casilla 19001, Santiago 19, Chile  \label{inst4}
\and
INAF-Osservatorio Astronomico di Capodimonte, Via Moiariello 16, I-80131 Naples, Italy  \label{inst5}
}

   \date{Received ; accepted ...}

 
  \abstract
{Globular  clusters associated with  the  Galactic  bulge are  important tracers  of  stellar  populations  in  the inner Galaxy. High resolution analysis of  stars in these clusters allows us to characterize them in terms of kinematics, metallicity, and individual abundances, and to compare these fingerprints  with those characterizing field populations.}
{ We present iron and element ratios for seven red giant
stars in the globular cluster NGC~6723, based on high resolution 
spectroscopy.}
 {High  resolution spectra ($\textmd{R}\sim48~000$)  of seven  K giants  belonging to  NGC 6723  were
obtained   with   the  FEROS   spectrograph   at   the  MPG/ESO   2.2m
telescope.  Photospheric parameters  were  derived  from  $\sim130$  \ion{Fe}{I}  and  \ion{Fe}{II}
transitions. Abundance ratios were obtained from line-to-line spectrum
synthesis calculations on clean selected features.}
{An intermediate  metallicity of [Fe/H]$=-0.98\pm0.08$ dex  and a  heliocentric radial
velocity of $v_{hel}=-96.6\pm1.3 \ \textmd{km} \ \textmd{s}^{-1}$ were
found for NGC 6723.  Alpha-element abundances present enhancements of
$\textmd{[O/Fe]}=0.29\pm0.18$ dex,  $\textmd{[Mg/Fe]}=0.23\pm0.10$ dex,  $\textmd{[Si/Fe]}=0.36\pm0.05$ dex,  and  $\textmd{[Ca/Fe]}=0.30\pm0.07$ dex.  Similar
overabundance is found for the  iron-peak Ti with $\textmd{[Ti/Fe]}=0.24\pm0.09$ dex.  Odd-Z
elements Na and Al present  abundances of $\textmd{[Na/Fe]}=0.00\pm0.21$ dex and $\textmd{[Al/Fe]}=0.31\pm0.21$ dex, 
respectively.  Finally,   the  s-element   Ba  is  also   enhanced  by
$\textmd{[Ba/Fe]}=0.22\pm0.21$ dex.}
{The enhancement levels of NGC 6723 are comparable to those of other metal-intermediate bulge globular clusters. In turn, these enhancement levels are compatible with the abundance profiles displayed by bulge field stars at that metallicity. This hints at a possible similar chemical evolution with globular clusters and the metal-poor of the bulge going through an early prompt chemical enrichment.}


   \maketitle

\section{Introduction}
\label{sec:introduccion}

The structure, age, and chemodynamical evolution of the Galactic bulge,
together with its relation with other Galactic components, is currently
debated.  Several studies    \citep{zoccali2008, babusiaux2010, 
hill2011, ness2013,   zoccali2014,  rojas_arriagada2014}, based on a few
hundred to several thousand stars, have constructed  a new picture
of  the bulge  as  an  extremely complex  structure, including at least
two stellar populations with different origins. In fact, bulge data shows chemodynamical distribution that are consistent with stars belonging to i) a narrow, metal-rich, boxy/peanut X-shaped component with bar-like kinematics, and ii) a broad metal-poor, kinematically hot population, which can be interpreted as a classical bulge coexisting with the Galactic bar.

The study of  bulge field stars is complicated by  the large spread in
age  and distance,  and by  the  contamination from  disk field  stars
overlapping  in the  lines of  sight. In addition, the  transition between
internal (thick) disk and  bulge is  still  not totally  understood. In  this
context,  the study  of  globular  clusters as  tracers  of the  bulge
populations is a  valuable tool because they provide  clean points in
the  age-metallicity-distance  distributions.  Even if  globular clusters  are  not
necessarily representative of the field stellar populations, they must
be included in the bulge formation scenario.

The   globular    cluster   NGC    6723   is   located    at   (J2000)
$\alpha=18^{\textmd{h}}59^{\textmd{m}}33.2^{\textmd{s}}$,
$\delta=-36^{\circ}37'54''$, projected  at about $-17^{\circ}$ from
the   Galactic   center   on   the   minor   axis   ($l=0.07^{\circ}$,
$b=-17.299^{\circ}$).    According  to   the  2010   edition  of   the
\cite{harris_catalogo}   catalog,   NGC~6723    is   rather   bright
($M_V=-7.83$ mag), fairly large in size ($r_t=10.5'$), and rather close to
the Galactic center ($R_{GC}=2.6\ \textmd{kpc}$). Its proximity  to the Galactic center makes  it fall in
the common definition of {\it bulge cluster} ($R_{\textmd{GC}}<5\ \textmd{kpc}$), although its  nearly  polar orbit  suggests membership  in the  inner  halo
\citep{dinescu}.

The most recent CMD of NGC 6723 available  in the literature
is by  \citet{lee2014}. An updated census of the variable  star content of NGC 6723 is performed from their high-precision $BV$ CCD photometry. Previous work in this direction goes back to the photographic  work by \citet{menzies}.  From a Fourier decomposition analysis of the RR Lyrae variables, \citet{lee2014} obtain [Fe/H]$=-1.23\pm0.11$ dex and $E(B-V)=0.063\pm0.015$ mag with a distance modulus of $(m-M)_0=14.65\pm0.05$ mag.

On the other hand, spectroscopic  measurements  characterize NGC  6723  as  an
intermediate metallicity cluster. A  high  metallicity  of  [Fe/H]=$-0.7$ dex  was  obtained  by
\citet{smith1981} from low  resolution (R=1300) spectra of  five RR Lyrae
stars   using  the   $\delta  S$   parameter.  \citet{rutledge1997}   obtained   [Fe/H]=$-1.09$ dex from   calcium  triplet
measurements,   while
\citet{kraft2003} derived a
metallicity  of  [Fe/H]=$-1.12$ dex by recalibrating  the \citet{rutledge1997} value. Additionally, two studies perform a chemical characterization of NGC
6723, based  on samples of high resolution spectra. The results of \citet{fullton96} 
come from the  analysis of a small sample of three RGB stars at $R\sim33000$,  deriving mean
values  of [Fe/H]=$-1.26$ dex,  [Ca/Fe]$=0.33$ dex,   [Ti/Fe]$=0.24$ dex,   and
[Si/Fe]$=0.68$ dex. Unfortunately, this  work has never been  published in a
refereed  journal,  so  details  of  the  analysis  are  not  publicly
available. The recent work of \cite{gratton2015} focus on the chemical analysis of a statistically significant sample of 30 red horizontal branch (RHB) and 17 blue horizontal branch (BHB) stars, observed at a maximum resolution of $R\sim18700$. From this sample, they inferred an average radial velocity of $V_{rad}=-95.8\pm0.6 \ \textmd{km} \ \textmd{s}^{-1}$ and mean values of $\textmd{[Fe/H]}=-1.22\pm0.08$ dex, $\textmd{[O/Fe]}=0.53\pm0.09$, $\textmd{[Na/Fe]}=0.13\pm0.09$, $\textmd{[Mg/Fe]}=0.51\pm0.06$, $\textmd{[Si/Fe]}=0.60\pm0.08$, $\textmd{[Ca/Fe]}=0.81\pm0.13,$ and $\textmd{[Ba/Fe]}=0.75\pm0.17$ dex for NGC~6723. Complementary high resolution spectroscopic studies, targeting globular cluster samples at different evolutionary stages, are important to accurately characterize their  chemical signatures  and shed light on their eventual complexities. This is especially important in that our current understanding of globular clusters  departs from the  classical hypothesis of simple stellar populations.

We carried out  an observing campaign at the 2.2m  telescope at ESO La
Silla Observatory, to  obtain time  series photometry with  the Wide  Field Imager
 \citep[WFI;][]{baade1999} and high  resolution spectroscopy with the
Fiber-fed    Extended     Range    Optical     Spectrograph    \citep[FEROS;][]{kaufer1999}. We  analyze spectra of eight  giants (from which seven turn out to be cluster candidates), deriving iron and element ratios. The  analysis of the CMD and variable star content  of the cluster will  be published in a forthcoming paper. 

 The paper is organized as follows:
 The  observations  are  described in the next section,  while
 initial  photometric   stellar  parameters  are  presented   in  Sect. \ref{sec:parametros_fotometricos}.  The adopted line list is described in Sect. \ref{sec:linelist}.  Iron abundances are  derived in Sect. \ref{sec:iron_abundance}. Sect. \ref{sec:extincion}   discusses distance and  a foreground interstellar reddening values  derived from  the spectroscopic  parameters. In Sect. \ref{sec:abundance_ratios} we present our derived abundance ratios. Discussion and conclusions are in Sect. \ref{sec:discusion_and_conclusions}.

\section{The data}
\label{sec:datos}

\begin{figure}
\begin{center}
\includegraphics[width=9.22cm]{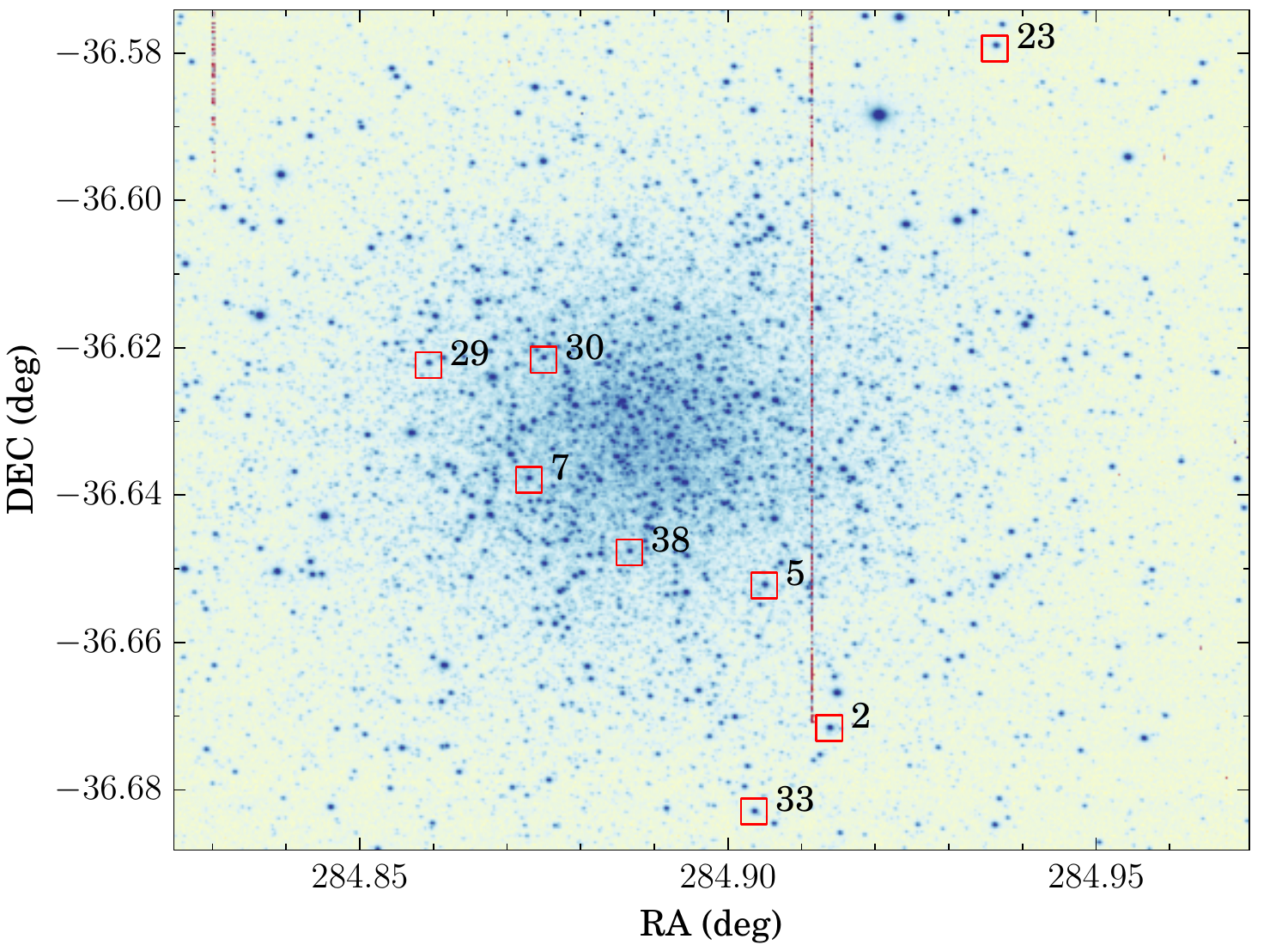}
\caption{ Finding chart showing the location of the 8 stars observed in NGC 6723.  The background image correspond to a $\sim8.76' \times 6.84'$ section of a WFI image in the BB\#V/89-ESO843 filter.}
\label{fig:finding_chart}
\end{center}
\end{figure}

\begin{figure}
\begin{center}
\includegraphics[width=9cm]{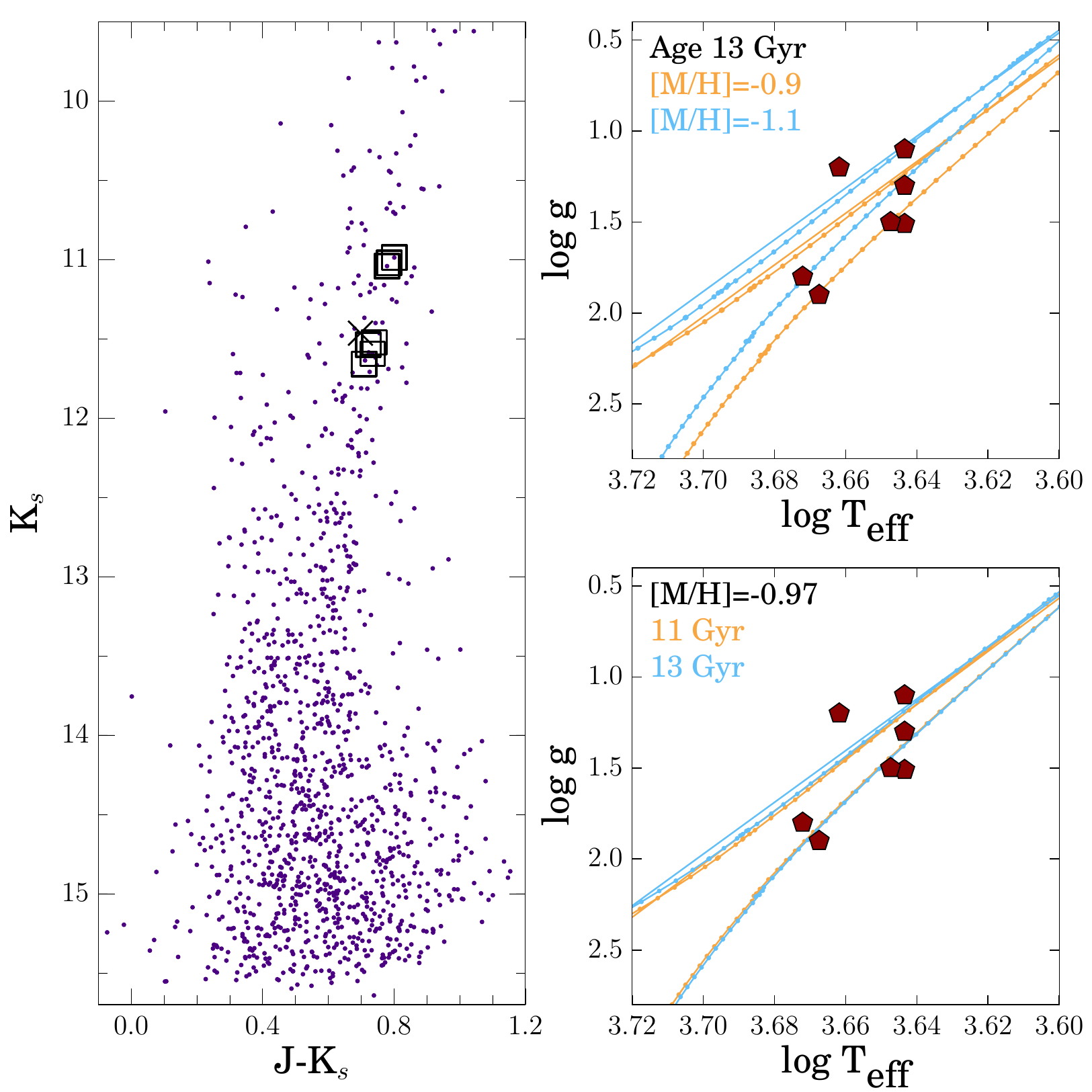}
\caption{ \textit{Left panel: }$K_s$  vs.
  $J-K_s$ 2MASS  based CMD of  the cluster. The  spectroscopic targets
  are marked with black empty squares. The field contaminant star is marked with a cross symbol. \textit{Right  panels:  } $\log(g)$ vs. T$_{\textmd{eff}}$
  diagrams for the spectroscopic sample.  PARSEC \citep{bressan2012} 13 Gyr isochrones for
  metallicity [M/H]$=-0.9$ dex and [M/H]$=-1.1$ dex are overplotted in the upper
  panel,  while two  isochrones  of 11  and 13  Gyr  with the  cluster
  metallicity  in  the  lower  panel.}
\label{fig:cmd_params}
\end{center}
\end{figure}

Spectra for eight stars in the globular cluster NGC 6723  were obtained
with  the FEROS  spectrograph \citep{kaufer1999}  at the  MPG/ESO 2.2m
telescope  (La  Silla  Observatory),  as part  of  the  ESO  program
085.D-0143. FEROS spectra are in the range $3500-9200\ \AA$, spread
over 39  echelle orders with a  resolution of  $\textmd{R}\sim48~000$. Two
fibers allowed us to obtain  a stellar spectrum  plus a  sky simultaneously
during a single exposure. The observed  sample stars, at the sky positions depicted in Fig. \ref{fig:finding_chart}, were selected from the K giant region of the  cluster CMD, as shown in the  left panel of Fig.
\ref{fig:cmd_params}. Table  \ref{tab:fotometria_targets} lists  the
coordinates, 2MASS IDs \citep{2mass_catalogo} and the  $VJHK_s$
photometry of the observed stars. $J$, $H,$ and $K_s$ magnitudes come from 2MASS, while $V$ corresponds to a visual magnitude from the UCAC3 catalog \citep{ucac_catalog}.

The spectra  were  flat-fielded,  optimally   extracted,  wavelength
calibrated,  and   corrected  to   heliocentric  system   of  reference
(barycentric  correction)  with  the standard    FEROS-DRS
pipeline\footnote{\url{http://www.eso.org/sci/facilities/lasilla/instruments/feros/tools/DRS.html}} based on MIDAS.
For most of the stars, the total exposure time was split into two or more exposures to remove cosmic rays.  A log  of observations is
listed  in   Table   \ref{tab:log_obs}.    We  use   the   IRAF\footnote{IRAF is distributed by the National Optical Astronomy Observatory, which is operated by the Association of Universities for Research
in Astronomy (AURA) under cooperative agreement with the National Science Foundation.} task
\textit{skytweak} to remove the sky emission lines from the individual
spectra with the corresponding companion  sky fiber. After applying the \textit{continuum} task and a cubic spline fit, we normalized the stellar continuum to the unit.

The radial  velocity of each  individual spectra was derived  by cross-correlation  (IRAF task  \textit{fxcor})  with a  synthetic  template. This template was  generated adopting stellar  parameters typical of  K giant
stars, \textbf{i.e.} $\textmd{T}_{\textmd{eff}}=4500   \    \textmd{K}$,
$\textmd{[Fe/H]}=-1.0  \ \textmd{dex}$,  $\log(g)=1.8 \ \textmd{dex,}$ and  $v_t=1.4\ \textmd{km}\ \textmd{s}^{-1}$ .
The corrected final rest-frame spectra were obtained  by applying the
velocity  correction  with  \textit{dopcor}.  Post-processed  spectra
corresponding to  individual exposures of  each star were  then median
combined.  A mean heliocentric radial velocity of $v_{hel}=-96.6\pm1.3
\ \textmd{km} \ \textmd{s}^{-1}$ was  found for NGC 6723, which is in excellent
agreement  with  the  value   of  $v_{hel}=-94.5\pm3.6  \  \textmd{km}
\     \textmd{s}^{-1}$    reported     in    the     compilation    by
\citet{harris_catalogo}. One  of the  program stars (\#8) is  not a
member  of  the  cluster  according  to  its  radial  velocity.   The contaminant nature of this star  is confirmed  by the  metallicity determinations presented  in Sect. \ref{sec:iron_abundance}. The propagation of internal errors (the velocity errors from the cross-correlation) of individual stars gives $\sigma_{internal}=0.62$. The r.m.s velocity scatter from cluster members correspond to 3.37 $\textmd{km} \ \textmd{s}^{-1}$. These values imply an internal velocity dispersion of 3.3 $\textmd{km} \ \textmd{s}^{-1}$ for NGC~6723. In comparison, \citet{gratton2015} quote a value of 4.3 $\textmd{km} \ \textmd{s}^{-1}$. 

Signal-to-noise (S/N) ratios  were  measured  in  the  final  coadded  spectra  at  several
wavelength regions  spanning 2-3  \AA.  The  values reported  in Table
\ref{tab:log_obs},    correspond     to    the    mean     of    these
measurements.\\

\begin{table*} 
\centering
\caption{Identifications, positions, and magnitudes of the target stars.}
\begin{tabular}{lccccccc}
\hline
\hline
Star ID & 2MASS ID         & $\alpha$(J2000.0) & $\delta$(J2000.0) & $J$      & $H$      & $K_s$      & $V_{\rm UCAC3}$      \\\hline
\#1      & 18593930-3640179 & 284.913763        & $-$36.67165         & 11.788 & 11.124 & 10.987 & 14.079 \\
\#2       & 18593718-3639082 & 284.904941        & $-$36.652302        & 11.806 & 11.163 & 11.020 & 14.048 \\
\#3      & 18592950-3638165 & 284.872956        & $-$36.637917        & 11.820 & 11.089 & 11.041 & 14.051 \\
\#4      & 18592624-3637204 & 284.859371        & $-$36.622349        & 12.263 & 11.611 & 11.522 & 14.375 \\
\#5      & 18592998-3637177 & 284.874931        & $-$36.621597        & 12.259 & 11.611 & 11.537 & 14.385 \\
\#6      & 18593684-3640585 & 284.903519        & $-$36.682938        & 12.329 & 11.696 & 11.594 & 14.407 \\
\#7      & 18593278-3638519 & 284.886611        & $-$36.647755        & 12.368 & 11.737 & 11.659 & 14.515 \\
\#8      & 18594469-3634457 & 284.936217        & $-$36.579371        & 12.162 & 11.397 & 11.464 & 14.524 \\\hline
\end{tabular}
\label{tab:fotometria_targets}
\end{table*}

\begin{table*} 
\centering
\caption{Log of spectroscopic observations. For stars with several individual exposures, the signal-to-noise value corresponds to that of the final averaged spectrum.}
\begin{tabular}{lcccccccc}
\hline
\hline
Target & Date       & UTC          & Exp  & Airmass & Seeing & (S/N)/px & $v_{hel}$ & Corr$_{hel}$  \\
       &            &              & (s)  &         &  (")   &          & km s$^{-1}$ & km s$^{-1}$ \\\hline
\#1      & 2010-05-05 & 09:49:21.971 & 2700 & 1.057   & 0.93   & 61       & $-$96.65  &  24.31   \\
       & 2010-05-06 & 08:10:16.920 & 5400 & 1.014   & 1.00   &          & $-$96.52  &  24.20   \\
\#2      & 2010-05-05 & 05:46:51.973 & 5400 & 1.160   & 1.22   & 54       & $-$90.01  &  24.67   \\
\#3      & 2010-05-05 & 07:30:43.969 & 5400 & 1.021   & 1.11   & 58       & $-$100.38 &  24.52   \\
\#4     & 2010-06-22 & 01:24:06.707 & 4600 & 1.450   & 1.47   & 37       & $-$99.78  &  6.14    \\
       & 2010-06-22 & 03:11:28.710 & 4600 & 1.103   & 1.77   &          & $-$99.82  &  6.01    \\
\#5     & 2010-06-22 & 04:33:50.710 & 4500 & 1.018   & 1.56   & 42       & $-$96.16  &  5.89    \\
       & 2010-06-22 & 05:50:02.710 & 4500 & 1.027   & 1.54   &          & $-$96.20  &  5.75    \\
\#6     & 2010-06-22 & 07:42:12.708 & 4500 & 1.211   & 1.47   & 26       & $-$94.08  &  5.57    \\
       & 2010-06-22 & 08:58:04.707 & 4500 & 1.540   & 1.62   &          & $-$94.25  &  5.46    \\
\#7     & 2010-06-23 & 04:18:17.468 & 4500 & 1.024   & 0.84   & 42       & $-$98.84  &  5.44    \\
       & 2010-06-25 & 05:06:03.250 & 3600 & 1.011   & 1.39   &          & $-$99.01  &  5.31    \\
       & 2010-06-25 & 06:06:55.251 & 3600 & 1.043   & 1.55   &          & $-$99.05  &  4.29    \\
\#8     & 2010-06-23 & 01:12:15.468 & 4500 & 1.500   & 1.07   & 31       &  7.29   &  5.71    \\
       & 2010-06-23 & 02:28:33.467 & 4500 & 1.191   & 0.91   &          &  7.35   &  5.63    \\       
       \hline
\end{tabular}
\label{tab:log_obs}
\end{table*}

\section{Photometric stellar parameters}
\label{sec:parametros_fotometricos}

We estimate  photometric atmospheric  parameters as  a first  guess to
perform a more precise  spectroscopic analysis. Effective temperatures
$\textmd{T}_{\textmd{eff}}$ were  derived from  the $V-J$, $V-H$,  $V-K,$ and
$J-K$      colors using      the     recent      calibrations  by
\citet{gonzalez-bonifacio_calib}. These  calibrations were constructed
by performing a fully  self-consistent infrared flux method  in the 2MASS
photometric  system, and  thus are  the best  choice available  in the
literature. To use them, we adopt a metallicity of [Fe/H]$=-1.0$ dex
and a  mean reddening  of $E(B-V)=0.05$ mag for NGC 6723, as taken  from the 2010 edition of the \citet{harris_catalogo} compilation.  We adopted the extinction law given by 
\citet{cardelli89}, i.e.,  $R_V=A_V/E(B-V)=3.1$, $E(V-J)/E(B-V)=2.26$,
$E(V-H)/E(B-V)=2.51$, $E(V-K)/E(B-V)=2.75,$       and
$E(J-K)/E(B-V)=0.52$.\\ The derived photometric effective temperatures
are      listed      on      the     left      side      of      Table
\ref{tab:parametros_fotom_y_espec}.\\             Using    those
$\textmd{T}_{\textmd{eff}}$  values,   we  derive  gravity   from  the
classical relation
$$\log g_{\star}=4.44+4\log\frac{T_{\star}}{T_\odot}+0.4\left(M_{bol\star}-M_{bol\odot}\right)+\log\frac{M_{\star}}{M_\odot},$$
adopting   $T_\odot=5770$ K,  $M_{\star}=0.85M_\odot$,
$M_{bol\odot}=4.75$, a  distance modulus  of $(m-M)_V=14.84$ mag for NGC~6723 (Harris compilation),  and the
reddening  already   used  to  derive the temperatures.   The  bolometric
corrections from \citet{alonso_calib} and  the estimated gravities are
given  in   the  last  two   columns  of   the  left  side   of  Table
\ref{tab:parametros_fotom_y_espec}.

\begin{table*}
\centering
\caption{Photometric temperatures derived from $J-K$, $V-J$, $V-H$ and $V-K$ colors, bolometric corrections, corresponding derived gravities and   final spectroscopic parameters of the target stars.}
\begin{threeparttable}
 \centering
\begin{tabular}{c|cccccc|cccccc}
\hline
\hline
& \multicolumn{5}{c}{\hbox{}} Photometric\phantom{-} parameters  & \multicolumn{5}{c}{\hbox{}}
 Spectroscopic\phantom{-} parameters\\
\cline{2-7}  \cline{7-13} 
\rule{0pt}{4ex}  Star & T($J-K$)\tnote{a} & T($V-J$)\tnote{b} & T($V-H$)\tnote{c} & T($V-K$)\tnote{d} & BC$_\textmd{v}$ &  $\log(g)$ & T$_{\textmd{eff}}$ & $\log(g$) & [\ion{Fe}{I}/H]& [\ion{Fe}{II}/H]  &  $v_t$  &   $V_{\textmd{hel}}$ \\
     &   K    &   K    &   K    &   K    &  mag       &  dex      &    K     & dex    & dex   & dex  &  km s$^{-1}$ & km s$^{-1}$  \\ \hline    
\#1    &  4359  & 4353   & 4261  &  4268  & $-$0.29 & 1.48 & 4400  &  1.51 &  $-$0.96 & $-$0.95 &  1.3 &  $-$96.586 \\  
\#2    &  4399  & 4397   & 4309  &  4310  & $-$0.27 & 1.49 & 4400  &  1.30 &  $-$1.05 & $-$1.04 &  1.3 &  $-$90.010 \\ 
\#3    &  4417  & 4407   & 4257  &  4322  & $-$0.27 & 1.49 & 4440  &  1.50 &  $-$0.99 & $-$1.05 &  1.4 & $-$100.377 \\
\#4   &  4522  & 4520   & 4395  &  4431  & $-$0.22 & 1.69 & 4700  &  1.80 &  $-$0.93 & $-$0.93 &  1.4 &  $-$99.787 \\ 
\#5   &  4576  & 4506   & 4388  &  4436  & $-$0.22 & 1.70 & 4650  &  1.90 &  $-$0.91 & $-$0.94 &  1.3 &  $-$95.182 \\ 
\#6   &  4539  & 4555   & 4435  &  4461  & $-$0.21 & 1.72 & 4400  &  0.60 &  $-$0.87 & $-$0.84 &  0.9 &  $-$94.164 \\ 
\#7   &  4614  & 4486   & 4385  &  4430  & $-$0.22 & 1.75 & 4400  &  1.10 &  $-$1.10 & $-$1.09 &  1.0 &  $-$98.966 \\
\#8   &  4646  & 4293   & 4150  &  4289  & $-$0.28 & 1.68 & 4500  &  1.90 &  $-$0.29 & $-$0.31 &  1.0 &    7.322 \\\hline
\end{tabular}
\begin{tablenotes}
  \item[a]  \scriptsize{Calibration standard deviation: 94 K.}
  \item[b]  Calibration standard deviation: 18 K.
  \item[c]  Calibration standard deviation: 23 K.
  \item[d]  Calibration standard deviation: 23 K.
\end{tablenotes}
\end{threeparttable}

\label{tab:parametros_fotom_y_espec}
\end{table*}

\section{The line list}
\label{sec:linelist}

The \ion{Fe}{I} line  list and  respective oscillator  strengths used
here  includes   $\sim130$  transitions,   and  are   described  in
\citet{zoccali_2008}.  Also  included in  the list are ten \ion{Fe}{II}
transitions carefully selected to have a minimum impact from blends in
K giants. In our analysis, we only used  a part of the reddest portion
of  the FEROS spectrum  from 5500  to 6800  \AA. At shorter wavelengths, line density is high and some molecular bands are present,  making  the continuum determination difficult.
Because of the small amount of available lines, \ion{the Fe}{II} constraint to [Fe/H] is sensitive to the error introduced by blended and small undefined lines.   To ensure  a clean  analysis, we  visually examined
each  spectrum   to  check  the  quality  of  the  \ion{Fe}{II}  lines,
eliminating  those lines that are too small or suspicious
blending  cases from the subsequent analysis.  In  the  analysis ,  we  used  only  \ion{Fe}{I}  and
\ion{Fe}{II}   lines    with   equivalent    widths   $\textmd{EW}<200
\ \textmd{m\AA}$.

Lines  and respective atomic constants for odd-Z elements Al, Na,  $\alpha$-elements Mg, Si,  Ca, Ti, and heavy
element  Ba  were  adopted
from \citet{zoccali2008}. The specific set of lines  was verified to be clean from
blends from telluric lines in  our target stars. The  damping constants
$C_6$ were adopted from \citet{coelho2005}. They were computed where possible, and in particular for most of the
\ion{Fe}{I}  lines,   using  the  collisional  broadening   theory  of
\citet{barklem1998,  barklem2000}.  In  order  to  check the  internal
consistency of our  chemical analysis with this line list,  we took a
solar spectrum from  the FEROS archive, reducing it in  the same way as
the  target  stars.   Spectrum  synthesis  calculations  were then used  to
reproduce the solar spectrum. In those  cases where the fit to the Sun
could  be  improved, the  oscillator  strength value $\log(gf)$ of  the
respective line  was changed until a  best fit was obtained.    We  adopted the  oscillator strength
$\log(gf)=-9.716$   derived   by    \citet{allende_prieto2001} for the
forbidden oxygen  line at  6300 \AA.    Solar
abundances were  adopted from \citet{grevesse_sauval}, except  for
oxygen, where a value of $\epsilon(O)=8.77$ was adopted, as recommended
by \citet{allende_prieto2001} for the use of 1D model atmospheres.

\section{Iron abundances}
\label{sec:iron_abundance}
Equivalent widths were measured in the coadded spectra via the automatic code
DAOSPEC \citep{daospec_paper}.

\begin{figure}
\begin{center}
\includegraphics[width=9cm]{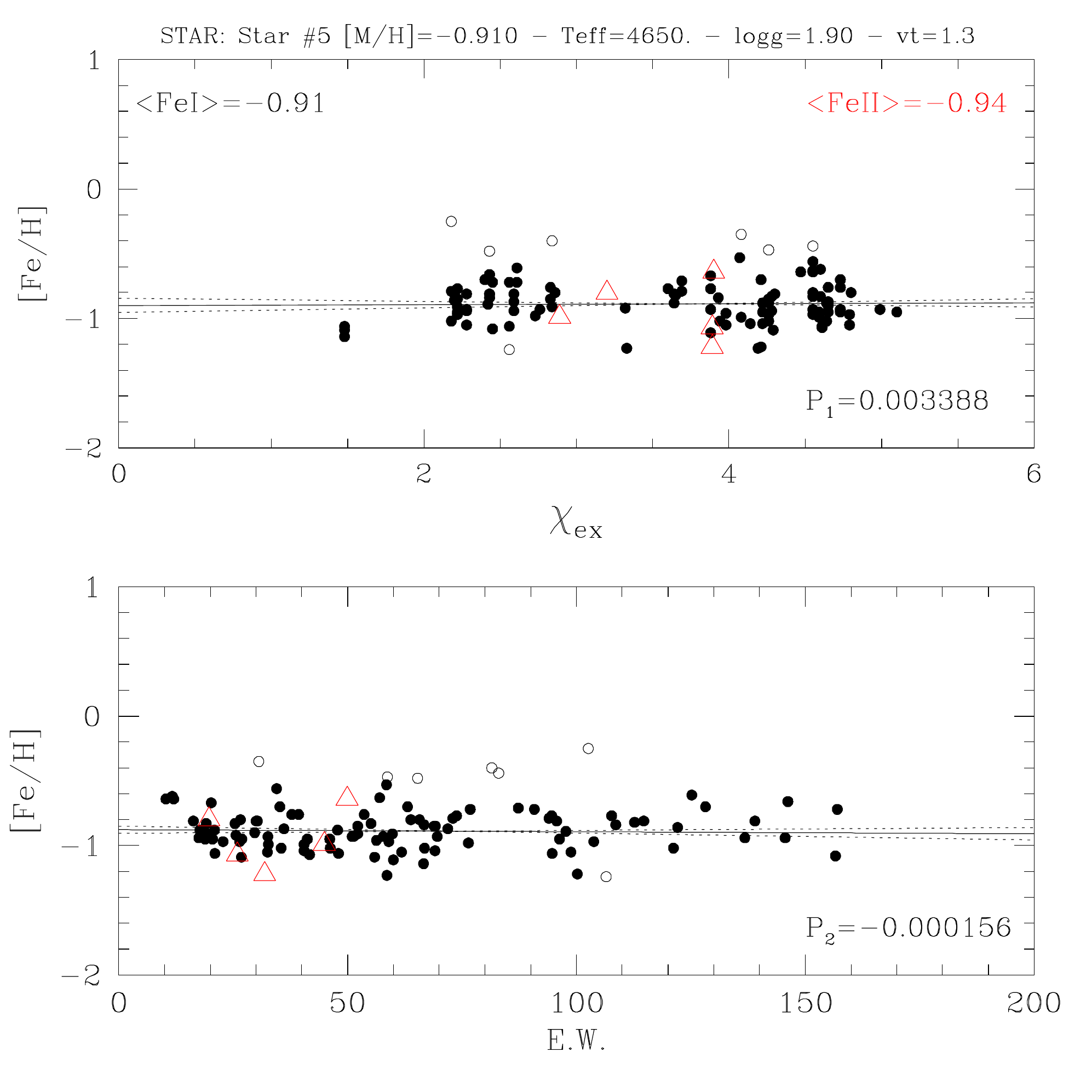}
\caption{\ion{Fe}{I} (black solid dots) and \ion{Fe}{II} (red open triangles) abundances versus $\chi_{ex}$ (eV) and EW$_p$ (m\AA) for star \#5. Open circles stand for \ion{Fe}{I} lines excluded from computation of the mean abundance using a $\sigma$ clipping criterion. Temperature and microturbulence are estimated by eliminating the trend of \ion{Fe}{I} abundance with respect to $\chi_{ex}$  and EW$_p$. } 
\label{fig:diag_diag}
\end{center}
\end{figure}

Model atmospheres were  interpolated from the MARCS  grid of spherical
models\footnote{\url{http://www.marcs.astro.uu.se/}}    described   in
\citet{marcs_2003}. A LTE analysis was  performed with an improved and
updated version of the classical code presented in \citet{spite_1967}, using the
atmospheric  models  and  EW measured  for  the  \ion{Fe}{I}  and
\ion{Fe}{II}  lines   in  the   list.  A   solar  iron   abundance  of
$\epsilon$(Fe)=7.50 \citep{grevesse_sauval}  was adopted.  The stellar
parameters were  derived through  an iterative process,  starting from
the photometric   temperature   and  gravity   as  an   initial
guess. Temperature was further  constrained by imposing the excitation
equilibrium of \ion{Fe}{I}  lines, namely,  that lines  of different
excitation  potential  $\chi_{ex}$ gave  the  same iron abundance.  The
measured \ion{Fe}{II} lines  enabled us to derive  gravity by imposing
ionization  equilibrium (\ion{Fe}{I}  and  \ion{Fe}{II} abundances  in
agreement).   Microturbulence  velocity   $v_t$  was  determined   by
eliminating the trend of \ion{Fe}{I}  abundances versus EW$_p$. We used
predicted  EW$_p$  instead  of  observed  EW   to  avoid  the introduction of systematics due to the correlation of errors in EW and
metallicity,  as explained  in  \citet{zoccali_2008}. Iron  abundances
were obtained  as the  weighted mean  of the  line-by-line measurements,
where the  weight associated with each  line is the inverse  square of
its  abundance   error,  as   derived  from  the   error  in   the  EW
measurement.    The   procedure    is   illustrated  for star  \#5  in
Fig.   \ref{fig:diag_diag}.   Final   spectroscopic   parameters
$\textmd{T}_{\textmd{eff}}$, $\log(g)$,   [Fe/H],  and $v_t$  are  given  in the  right  side of  Table
\ref{tab:parametros_fotom_y_espec}. The  star final iron  abundance [Fe/H] is
taken as the average between values estimated from \ion{Fe}{I}  and \ion{Fe}{II} lines.

It  is clear  from Table  \ref{tab:parametros_fotom_y_espec} that  star \#8, previously  discarded as a member of  the cluster because of its radial  velocity, has also a metallicity significantly
higher than the remaining  sample stars. Having confirmed its contaminant nature, we exclude it from the following analysis.

\begin{table} 
\centering
\caption{ \ion{Fe}{I} and \ion{Fe}{II} uncertainties induced by a change of $\Delta \textmd{T}_{\textmd{eff}}=100\ \textmd{K}$, $\Delta \log(g)=0.2$ dex, $\Delta v_t=0.2 \ \textmd{km} \ \textmd{s}^{-1}$, and corresponding total error.}
\begin{tabular}{llcccc}
\hline
\hline
Star & Specie &  $\Delta$T &  $\Delta \log(g)$ & $\Delta v_t$ & $\left(\sum x^2\right)^{1/2}$ \\\hline
\#1    & [Fe/H] (I) & $-$0.05 & $-$0.03 & 0.09 & 0.11  \\
     & [Fe/H] (II)&  0.13 & $-$0.11 &  0.05 & 0.18 \\
\#2    & [Fe/H] (I) &  0.06 &  0.03 & $-$0.10 & 0.12 \\
     & [Fe/H] (II)& $-$0.13 &  0.10 & $-$0.06 & 0.18 \\
\#3    & [Fe/H] (I) &  0.04 &  0.02 & $-$0.09 & 0.10 \\
     & [Fe/H] (II)& $-$0.08 & $-$0.16 &  0.00 & 0.18 \\
\#4   & [Fe/H] (I) &  0.09 &  0.03 & $-$0.08 & 0.12 \\
     & [Fe/H] (II)& $-$0.07 &  0.10 & $-$0.05 & 0.13 \\
\#5   & [Fe/H] (I) &  0.09 &  0.02 & $-$0.08 & 0.12 \\
     & [Fe/H] (II)& $-$0.10 &  0.10 & $-$0.06 & 0.15 \\
\#6   & [Fe/H] (I) &  0.07 &  0.02 & $-$0.18 & 0.19 \\
     & [Fe/H] (II)& $-$0.12 &  0.10 & $-$0.10 & 0.19 \\
\#7   & [Fe/H] (I) &  0.11 &  0.10 & $-$0.06 & 0.16 \\
     & [Fe/H] (II)&  0.11 &  0.35 &  0.17 & 0.40 \\
\#8   & [Fe/H] (I) &  0.03 &  0.03 & $-$0.15 & 0.16 \\
     & [Fe/H] (II)& $-$0.18 &  0.12 & $-$0.09 & 0.23 \\\hline
\end{tabular}
\label{tab:errores_espec}
\end{table}  

An average metallicity of [Fe/H]$=-0.98\pm0.08$ dex is derived for NGC 6723.

We estimate that    an   uncertainty   of   $\pm100$    K   in
$\textmd{T}_{\textmd{eff}}$ is well  inside the error in  the slope of
the  [Fe/H] vs. $\chi_{ex}$  trend used to constrain the temperature (Fig. \ref{fig:diag_diag}). Similarly, by imposing a flat trend of \ion{Fe}{I} line abundances as a function of EW$_p$, we can constrain $v_t$ with a precision of $\pm0.2 \textmd{km}\  \textmd{s}^{-1}$. Ionization equilibrium allows us to constrain $\log(g)$ within $\pm0.2$ dex.   We   estimate   errors   in metallicity from  \ion{Fe}{I}   and
\ion{Fe}{II}  lines   allowing   a  variation   of   $\Delta
\textmd{T}_{\textmd{eff}}=100$ K, $\Delta \log(g)=0.2$ dex, and $\Delta
v_t=0.2 \ \textmd{km}\  \textmd{s}^{-1}$, and quantifying their  impact in
the  final value.  The  induced error  corresponds  to the  difference
between an  abundance derived with  the altered  and  nominal
models.   Individual  errors   are  listed  in  columns   3-5  of  Table
\ref{tab:errores_espec} and the total error is listed in column 6. Average errors
associated with  \ion{Fe}{I}   and  \ion{Fe}{II} lines  are  0.14   and  0.21  dex,
respectively. These error bars are larger than the statistical uncertainty of 0.08 in the average cluster metallicity derived from the star individual values. This means that  our assumed uncertainties on $\textmd{T}_{\textmd{eff}}$, $\log(g)$ and $v_t$ are, in fact, probably overestimated given the good data quality of our spectra. In view of that, we adopt the statistical uncertainty of 0.08 dex as the error bar associated with the cluster metallicity estimate.

\section{Isochrone fitting: Distance and reddening}
\label{sec:extincion}

Since they are cluster members, our target stars are located essentially at the same distance  and are expected to have similar reddening values aside from small differential extinction across the cluster area. We use our derived spectroscopic parameters and available photometry to compute distances for our sample via an isochrone fitting technique. The algorithm (Rojas-Arriagada et al, \textit{in preparation}) makes use of the PARSEC set of isochrones to compute given atmospheric parameters and metallicity, a theoretical absolute magnitude $M_\lambda$ in a specific photometric band. Given $\textmd{T}_{\textmd{eff}}$, $\log(g)$, metallicity and corresponding errors, the algorithm selects a set of isochrones of all ages matching metallicity values inside the error bar.  A Monte Carlo sampling is generated in the $\textmd{T}_{\textmd{eff}}$-$\log(g)$ plane by considering the ellipsoid given by the respective errors. The distribution is projected onto the isochrones via closest match with a distance normalized by the errors as a metric. From the best match of each Monte Carlo sample, we obtain a theoretical absolute magnitude in the photometric band of interest. From this set of values, weighted by matching distance and evolutionary speed\footnote{This quantity is related to the mass differences $\Delta m$  between successive points in the isochrone, which gives in fact a proxy for the evolutionary speed of model stars at different evolutionary stages.}, a probability distribution function (\textit{pdf}) for $M_\lambda$ is constructed. A final absolute magnitude value for the star under analysis can be obtained as the mode or median of such distribution. Additionally, if we have photometry in $J$ and $K$ bands, a reddening estimate can be computed simultaneously with the distance. From the computed absolute magnitude $M_\lambda$, a distance modulus $\mu$, and then a distance is calculated. The distance modulus is corrected by reddening either using an input value or the value computed in-situ. Tests have shown that no significant differences arise from this choice.

\begin{figure}
\begin{center}
\includegraphics[width=9cm]{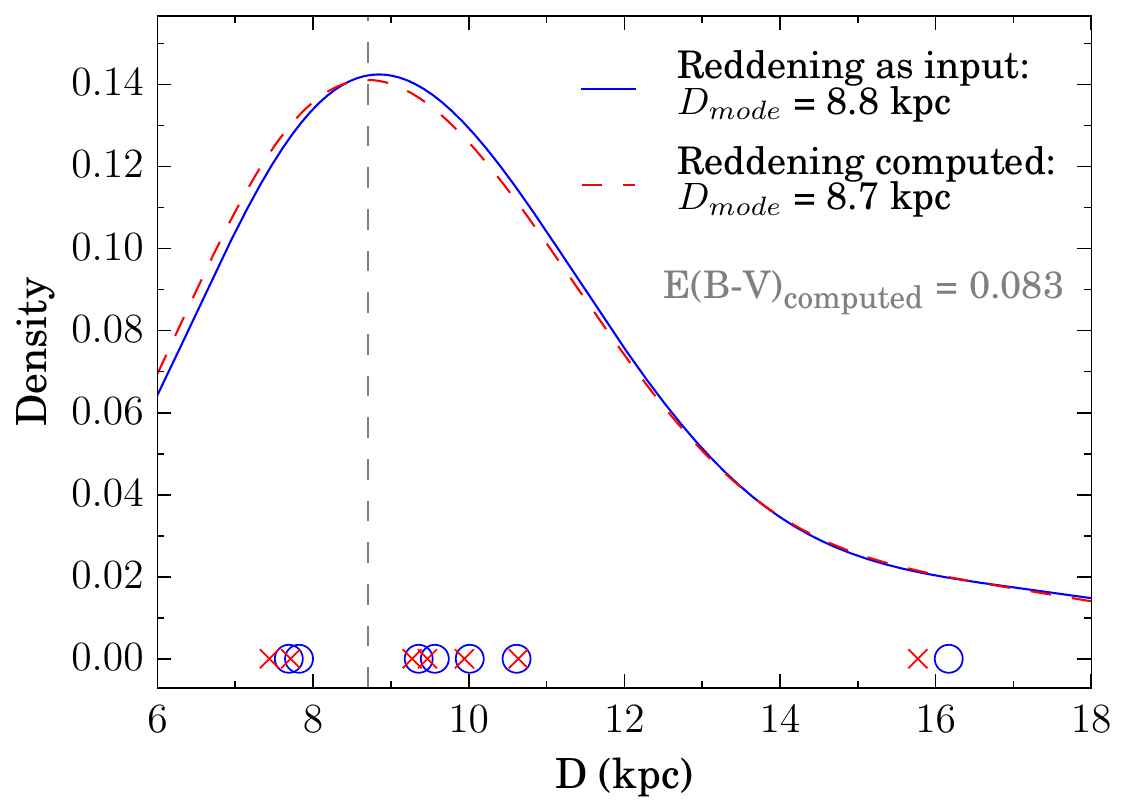}
\caption{Distance probability function of NGC~6723. Symbols at the bottom stand for the distances determined for individual stars. A blue solid line depicts the global \textit{pdf} generated from individual distances (depicted as blue circles) determined using reddening values from external measurements. On the other hand, a red dashed line depicts the global \textit{pdf} generated from individual distances (depicted by the red crosses) determined using reddening values computed simultaneously with the respective distances. In each case, errors on individual distance measurements are considered to compute the global cluster distance \textit{pdf}, assuming each individual star as a Gaussian random variable around the measured value. The modes of the  \textit{pdf}s are quoted, and provide our best estimates for the distance of NGC~6723. The estimated average reddening toward the cluster is also quoted in the panel. A vertical gray dashed line indicates the distance of NGC~6723 as given in the \citet{harris_catalogo} compilation.} 
\label{fig:distro_D}
\end{center}
\end{figure}

Using the procedure described above, we compute distances and foreground interstellar reddening for all the stars in the sample. The adopted errors in atmospheric parameters are the nominal parameters: $\Delta \textmd{T}_{\textmd{eff}}=100$ K, $\Delta \log(g)=0.2$ dex, and $\Delta \textmd{[Fe/H]}=0.15$ dex. We include the nonmember star \#8  as a consistency test; its resulting distance is the smallest of the sample, confirming its foreground contaminant nature. For each star, we have a set of four distance estimates because of the two 2MASS photometric bands employed ($J$ and $K_s$) and  the two ways to include reddening (external or estimated in-situ). In each case, we average results coming from $J$ and $K_s$ passbands as they have in fact a very small dispersion ($\sigma\sim0.22$ kpc in both cases). The two sets of final individual star distances are depicted by blue circles (determined using external reddening figures) and red crosses (determined using reddening calculated in-situ)  at the bottom of Fig. \ref{fig:distro_D}. Given the error figures estimated by the code for each distance, we can consider each star as a Gaussian random variable and then compute the global cluster distance \textit{pdf} as the superposition of individual contributions. The result of this is shown by the blue (external reddening) and red (reddening computed in-situ) curves. We can see that the selected method to include reddening in our calculations has, as anticipated, a low impact on the final cluster distance distribution. A gray dashed vertical line indicates the distance of NGC~6723 quoted in the \citet{harris_catalogo} compilation (8.7 kpc). We take our estimation of the cluster distance as the mode of the global \textit{pdf}. Our results are in excellent agreement with literature. Our final estimates are quoted in Figure \ref{fig:distro_D}. To be consistent, we adopt  our final estimate of the NGC~6723 distance as that coming from the analysis using reddening computed in situ. In this way we provide a distance of $D=8.7$ kpc and a reddening of $E(B-V)=0.083$ mag  for NGC~6723. The reddening is slightly larger than the $E(B-V)=0.05$ quoted in \citet{harris_catalogo}.

\section{Abundance ratios}
\label{sec:abundance_ratios}

Abundance ratios were estimated through line-by-line spectrum synthesis calculations for the lines listed in Table  \ref{tab:tablines}. For each line, synthesis calculations were carried with MOOG\footnote{Available at \url{http://www.as.utexas.edu/~chris/moog.html}.}, changing the respective elemental abundance until agreement with the observed line was reached. In general, errors in the abundance ratios can be attributed to the adopted $\log (gf)$ and damping constant C$_6$ while constructing the atomic line list. Some tests carried over selected lines, show that changing the $\log (gf)$ value by 0.1 dex produces an abundance variation of the same order for a fixed damping constant. 

In this analysis, molecular lines of CN  (A$^2\Pi$-X$^2\Sigma$), C$_2$ Swan (A$^3\Pi$-X$^3\Pi$), TiO (A$^3\Phi$-X$^3\Delta$) $\gamma,$ and  TiO (B$^3\Pi$-X$^3\Delta$) $\gamma$' systems are taken into account.
At the typical $\textmd{T}_{\textmd{eff}}$ of our stars, the main effect of molecule formation in their atmospheres is a lowering of the continuum level. By including molecular features in the spectrum synthesis, we improve the match with observations, alleviating systematics coming from bad continuum definition. 

\begin{figure}
\begin{center}
\includegraphics[width=8.8cm]{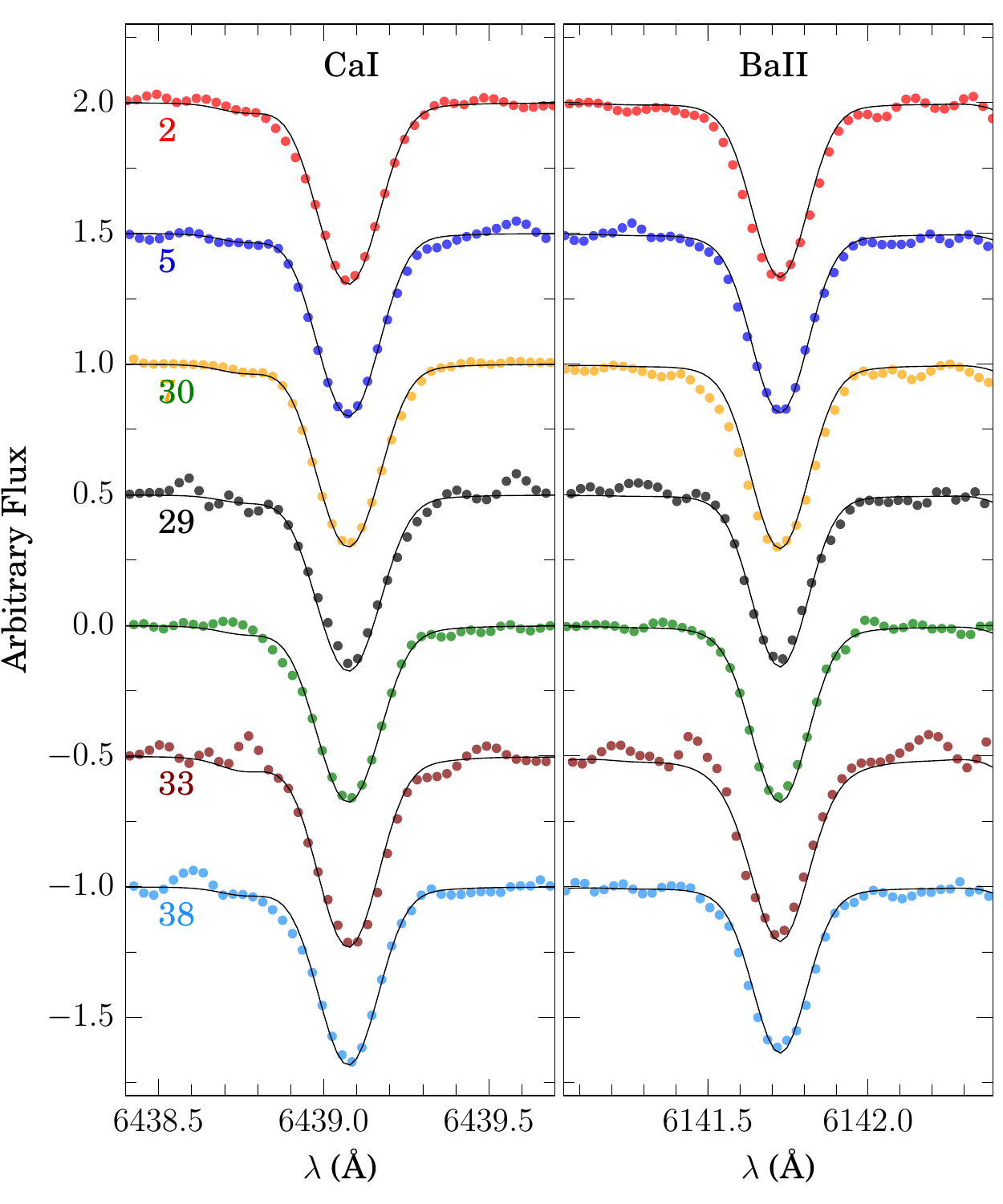}
\caption{\ion{Ca}{I} 6439.08 $\AA$  and \ion{Ba}{II} 6141.73 $\AA$ lines in the 7 cluster member stars. Observed spectra are depicted by color dots. A solid black line represents the best fit in each case.} 
\label{fig:sintesis_example}
\end{center}
\end{figure}

In Fig. \ref{fig:sintesis_example} we show the fits in the seven sample stars to the \ion{Ca}{I} 6439.08 $\AA$ and \ion{Ba}{II} 6141.73 $\AA$ lines. These examples illustrate the general good quality of the fits.
Line-by-line abundance ratios are presented for the sample stars in the last seven columns of Table \ref{tab:tablines}. Final elemental abundances are computed from the average of individual line values. They are listed in Table \ref{tab:abundancias_final} for each star. The final global cluster abundances are in the last row. 

As can be seen from Table \ref{tab:abundancias_final}, [Na/Fe] abundances show larger spread when compared with the other measured elements. Sodium abundances were determined from two lines at 6154.23 and 6160.75 $\AA$ (see Table \ref{tab:tablines}). During the process of spectrum synthesis for line fitting, we found that these lines are in general small in our sample  and not always well defined with respect to the noisy continuum. In fact, the two stars that present better quality are  star \#3, which has a combination of high S/N ratio and good values of seeing and airmass, and  star \#8, which is the noncluster foreground contaminant. This general modest quality for Na features can explain the larger error bar associated with our  final cluster [Na/Fe] estimate.

\begin{table*}
\caption{Abundance ratios [X/Fe] and line atomic parameters adopted.}
\begin{flushleft}
\begin{tabular}{cccccccccccc}
\noalign{\smallskip}
\hline
\noalign{\smallskip}
Species & ${\rm \lambda}$     & ${\rm \chi_{ex}}$ &  C$_6$ & $\log gf$ & \multispan7 [X/Fe]    \\ 
\cline{6-12}  \ 
  & (${\rm \AA}$) & (eV)  & (cm$^6$s$^{-1}$) &  & {\bf\#1}  & {\bf\#2} & {\bf\#3} & {\bf\#4} & {\bf\#5} & {\bf\#6} & {\bf\#7} \\ 
\noalign{\smallskip} 
\hline 
\noalign{\smallskip} 
\ion{O}{I} &  6300.310 &  0.00 & 3.00E$-$32 &   $-$9.72 & 0.40 & 0.25 & 0.00 & 0.42 & 0.25 & $-$0.05 & 0.15 \\ 
\ion{O}{I} &  6363.790 &  4.26 & 3.00E$-$32 &   $-$2.31 & 0.45 & 0.50 & 0.30 &  --- & 0.58 &  --- & 0.40 \\ 
\ion{Na}{I} &  6154.230 &  0.02 & 5.85E$-$34 &  $-$10.25 & $-$0.15 & $-$0.18 & 0.41 & $-$0.20 & 0.16 &  --- & 0.05 \\ 
\ion{Na}{I} &  6160.753 &  2.10 & 3.55E$-$31 &   $-$1.59 & $-$0.20 & $-$0.15 & 0.37 &  --- & 0.14 & $-$0.02 &  --- \\ 
\ion{Al}{I} &  6696.020 &  2.10 & 3.55E$-$31 &   $-$1.27 & 0.12 & 0.16 & 0.50 & 0.20 &  --- &  --- & 0.25 \\ 
\ion{Al}{I} &  6698.670 &  3.14 & 6.68E$-$31 &   $-$1.61 & 0.32 & 0.18 &  --- & 0.40 &  --- &  --- & 0.42 \\ 
\ion{Mg}{I} &  6318.720 &  3.14 & 6.67E$-$31 &   $-$1.95 & 0.25 & 0.20 & 0.25 & 0.23 & 0.25 & 0.22 &  --- \\ 
\ion{Mg}{I} &  6765.450 &  5.11 & 3.00E$-$31 &   $-$1.96 &  --- &  --- &  --- &  --- &  --- &  --- &  --- \\ 
\ion{Si}{I} &  6142.494 &  5.75 & 3.00E$-$32 &   $-$1.94 & 0.45 &  --- & 0.40 & 0.30 & 0.44 &  --- & 0.35 \\ 
\ion{Si}{I} &  6145.020 &  5.62 & 3.00E$-$32 &   $-$1.50 & 0.50 & 0.45 & 0.40 & 0.35 & 0.32 & 0.22 & 0.37 \\ 
\ion{Si}{I} &  6155.142 &  5.61 & 3.00E$-$32 &   $-$1.48 & 0.30 & 0.42 & 0.38 & 0.32 & 0.32 & 0.38 & 0.26 \\ 
\ion{Si}{I} &  6237.328 &  5.62 & 3.00E$-$31 &   $-$0.93 &  --- &  --- & 0.38 & 0.32 & 0.26 &  --- &  --- \\ 
\ion{Si}{I} &  6243.823 &  5.61 & 3.00E$-$31 &   $-$1.19 & 0.36 &  --- &  --- &  --- &  --- & 0.38 & 0.28 \\ 
\ion{Si}{I} &  6721.844 &  5.61 & 1.21E$-$30 &   $-$1.38 & 0.31 & 0.40 & 0.32 &  --- &  --- &  --- &  --- \\ 
\ion{Ca}{I} &  6156.030 &  5.86 & 1.89E$-$30 &   $-$1.20 & 0.31 &  --- &  --- &  --- &  --- &  --- &  --- \\ 
\ion{Ca}{I} &  6161.295 &  2.52 & 5.98E$-$31 &   $-$2.59 & 0.32 & 0.38 & 0.32 & 0.32 & 0.30 & 0.20 & 0.28 \\ 
\ion{Ca}{I} &  6162.167 &  2.52 & 5.00E$-$32 &   $-$1.29 & 0.25 & 0.25 & 0.27 & 0.38 & 0.30 & 0.35 & 0.32 \\ 
\ion{Ca}{I} &  6166.440 &  1.90 & 2.72E$-$31 &   $-$0.08 & 0.15 & 0.25 & 0.25 & 0.38 & 0.48 & 0.30 & 0.28 \\ 
\ion{Ca}{I} &  6169.044 &  2.52 & 3.57E$-$31 &   $-$1.14 & 0.20 & 0.28 & 0.28 & 0.50 & 0.40 & 0.30 & 0.40 \\ 
\ion{Ca}{I} &  6169.564 &  2.52 & 3.57E$-$31 &   $-$0.77 & 0.20 & 0.30 & 0.22 & 0.42 & 0.40 & 0.45 & 0.44 \\ 
\ion{Ca}{I} &  6439.080 &  2.53 & 2.39E$-$31 &   $-$0.44 & 0.20 & 0.30 & 0.21 & 0.42 & 0.40 & 0.55 & 0.45 \\ 
\ion{Ca}{I} &  6455.605 &  2.53 & 5.12E$-$32 &    0.30 & 0.20 & 0.20 & 0.23 & 0.33 & 0.33 & 0.25 & 0.30 \\ 
\ion{Ca}{I} &  6464.679 &  2.52 & 5.09E$-$32 &   $-$1.44 &  --- &  --- &  --- &  --- &  --- &  --- &  --- \\ 
\ion{Ca}{I} &  6471.668 &  2.52 & 5.06E$-$31 &   $-$2.42 & 0.20 & 0.35 & 0.23 & 0.30 & 0.38 & 0.25 & 0.40 \\ 
\ion{Ca}{I} &  6493.788 &  2.53 & 5.09E$-$32 &   $-$0.71 & 0.15 & 0.10 & 0.15 & 0.30 & 0.30 & 0.10 & 0.27 \\ 
\ion{Ca}{I} &  6499.654 &  2.52 & 1.82E$-$32 &    0.01 & 0.20 & 0.16 & 0.18 & 0.33 & 0.35 & 0.00 & 0.30 \\ 
\ion{Ca}{I} &  6508.846 &  2.52 & 5.05E$-$32 &   $-$0.82 &  --- &  --- &  --- &  --- &  --- &  --- &  --- \\ 
\ion{Ca}{I} &  6572.779 &  2.53 & 5.05E$-$32 &   $-$2.50 & 0.15 & 0.16 & 0.18 &  --- & 0.35 & 0.20 & 0.26 \\ 
\ion{Ca}{I} &  6717.687 &  0.00 & 2.62E$-$32 &   $-$4.39 & 0.25 & 0.30 & 0.38 & 0.40 & 0.35 & 0.50 & 0.57 \\ 
\ion{Ti}{I} &  6126.224 &  2.71 & 6.19E$-$31 &   $-$0.58 & 0.22 & 0.14 & 0.23 & 0.30 & 0.39 & 0.30 & 0.09 \\ 
\ion{Ti}{I} &  6258.110 &  1.07 & 2.06E$-$32 &   $-$1.42 & 0.33 & 0.16 & 0.20 & 0.33 & 0.38 & 0.40 & 0.09 \\ 
\ion{Ti}{I} &  6261.106 &  1.44 & 2.85E$-$32 &   $-$0.53 & 0.24 & 0.09 & 0.23 & 0.33 & 0.30 & 0.20 & 0.09 \\ 
\ion{Ti}{I} &  6303.767 &  1.43 & 4.68E$-$32 &   $-$0.58 & 0.20 &  --- & 0.21 & 0.45 & 0.35 &  --- &  --- \\ 
\ion{Ti}{I} &  6312.238 &  1.44 & 4.68E$-$32 &   $-$1.57 & 0.20 & 0.28 & 0.20 & 0.45 & 0.30 & 0.20 &  --- \\ 
\ion{Ti}{I} &  6336.113 &  1.46 & 4.75E$-$32 &   $-$1.55 & 0.20 & 0.19 &  --- & 0.45 &  --- &  --- &  --- \\ 
\ion{Ti}{I} &  6508.154 &  1.44 & 2.79E$-$32 &   $-$1.74 &  --- &  --- &  --- &  --- &  --- &  --- &  --- \\ 
\ion{Ti}{I} &  6554.238 &  5.59 & 3.00E$-$32 &   $-$3.50 & 0.17 & 0.10 & 0.14 &  --- & 0.30 & 0.27 & 0.25 \\ 
\ion{Ti}{I} &  6556.077 &  1.44 & 2.72E$-$32 &   $-$1.25 & 0.32 & 0.11 & 0.32 &  --- &  --- &  --- &  --- \\ 
\ion{Ti}{I} &  6599.113 &  1.46 & 2.74E$-$32 &   $-$1.23 & 0.20 & 0.05 & 0.28 & 0.15 & 0.35 &  --- & 0.20 \\ 
\ion{Ti}{I} &  6743.127 &  0.90 & 2.94E$-$32 &   $-$2.08 & 0.15 & 0.10 & 0.10 & 0.29 & 0.30 & 0.27 & 0.06 \\ 
\ion{Ti}{II} &  6491.580 &  0.90 & 2.89E$-$32 &   $-$1.81 & 0.35 &  --- & 0.30 & 0.45 & 0.30 & 0.27 &  --- \\ 
\ion{Ti}{II} &  6559.576 &  3.19 & 3.00E$-$32 &   $-$3.82 &  --- &  --- &  --- &  --- &  --- &  --- &  --- \\ 
\ion{Ti}{II} &  6606.970 &  2.06 & 9.86E$-$33 &   $-$2.18 &  --- &  --- &  --- &  --- &  --- &  --- &  --- \\ 
\ion{Ba}{II} &  6141.727 &  2.05 & 3.00E$-$32 &   $-$2.60 & 0.10 & 0.20 & 0.25 & 0.10 & 0.30 & 0.40 & 0.20 \\ 
\ion{Ba}{II} &  6496.908 &  2.06 & 9.63E$-$33 &   $-$2.99 & 0.00 & 0.20 & 0.26 & 0.10 & 0.30 & 0.40 & 0.20 \\ 
\noalign{\smallskip} \hline \end{tabular} 
\label{tab:tablines} 
\end{flushleft} 
\end{table*}

\begin{table*}
\begin{flushleft}
\caption{Final abundance ratios for cluster member stars. Global values and internal errors (dispersion around the mean) are in the last row.}
\label{tab:abundancias_final}      
\centering          
\begin{tabular}{ccccccccc} 
\noalign{\smallskip}
\hline\hline    
\noalign{\smallskip}
\noalign{\vskip 0.1cm} 
Star & [O/Fe] & [Na/Fe] & [Mg/Fe] & [Al/Fe] & [Si/Fe] & [Ca/Fe] & [Ti/Fe] & [Ba/Fe] \\
\noalign{\vskip 0.1cm}
\noalign{\hrule\vskip 0.1cm}
\noalign{\vskip 0.1cm}

\#1  & 0.43 & $-$0.18  & 0.25 & 0.22 &  0.38 & 0.21 & 0.22 &  0.05 \\
\#2  & 0.38 & $-$0.17  & 0.20 & 0.17 &  0.44 & 0.25 & 0.14 &  0.20 \\
\#3  & 0.15 &  0.39  & 0.25 & 0.50 &  0.39 & 0.23 & 0.21 &  0.26 \\
\#4 & 0.42  & $-$0.20  & 0.23 & 0.30 &  0.32 & 0.37 & 0.36 &  0.10 \\
\#5 & 0.42  &  0.15  & 0.25 &  --- &  0.34 & 0.36 & 0.33 &  0.30 \\
\#6 & $-$0.05 & $-$0.02  & 0.22 &  --- &  0.33 & 0.29 & 0.27 &  0.40 \\
\#7 & 0.28  &  0.05  & ---  & 0.34 &  0.32 & 0.36 & 0.13 &  0.20 \\
Mean & 0.29 $\pm$ 0.18 & 0.00$\pm$0.21 & 0.23$\pm$0.10 & 0.31$\pm$0.21 & 0.36$\pm$0.05 & 0.30$\pm$0.07 & 0.24$\pm$0.09 & 0.22$\pm$0.12 \\
\hline
\end{tabular}
\end{flushleft}
\end{table*}

\begin{table}[H]
\centering
\caption{Abundance uncertainties induced by a change of $\textmd{T}_{\textmd{eff}}=100\ \textmd{K}$, $\Delta \log(g)=0.2$ dex, $\Delta v_t=0.2 \ \textmd{km} \ \textmd{s}^{-1}$, and corresponding total error.}
\begin{tabular}{lcccc}
\hline
\hline
Abundance &  $\Delta$T &  $\Delta \log(g)$ & $\Delta v_t$ & $\left(\sum x^2\right)^{1/2}$ \\\hline
$\textmd{[O/Fe]}$   &  0.05 &  0.07 &  0.01 & 0.09 \\
$\textmd{[Na/Fe]}$  &  0.07 & $-$0.01 & $-$0.02 & 0.07 \\
$\textmd{[Mg/Fe]}$  &  0.03 &  0.00 &  0.00 & 0.03 \\
$\textmd{[Si/Fe]}$  & $-$0.04 &  0.00 & $-$0.10 & 0.11 \\
$\textmd{[Ca/Fe]}$  &  0.05 & $-$0.03 & $-$0.07 & 0.09 \\
$\textmd{[Ti/Fe]}$  &  0.13 &  0.01 & $-$0.04 & 0.14 \\
$\textmd{[Ba/Fe]}$  &  0.06 &  0.07 & $-$0.12 & 0.15 \\
$\textmd{[La/Fe]}$  &  0.07 &  0.14 &  0.01 & 0.16 \\\hline
\end{tabular}
\label{tab:errores_abund}
\end{table}

\section{Discussion and conclusions}
\label{sec:discusion_and_conclusions}

We present a high resolution analysis of red giant stars in the bulge globular cluster NGC~6723. Eight targets were originally selected from the K giant region of the cluster CMD. One of these targets turns out to be a foreground field contaminant, based on its discrepant radial velocity, metallicity, and derived distance. Using the clean cluster sample of seven stars, we derived a metallicity of [Fe/H]$=-0.98\pm0.08$ dex and a heliocentric radial velocity of $v_{hel}=-96.6\pm1.3 \ \textmd{km} \ \textmd{s}^{-1}$ for NGC~6723. Final mean abundance ratios for eight elements are reported in Table \ref{tab:abundancias_final}.

The odd-Z element sodium shows a solar ratio $\textmd{[Na/Fe]}=0.00\pm0.21$. From their horizontal branch-based study, \citet{gratton2015} report a somewhat larger value of $0.13\pm0.09$ dex.

The $\alpha$-elements oxygen, magnesium, silicon, and calcium are enhanced by $\textmd{[O/Fe]}=0.29\pm0.18$ dex, $\textmd{[Mg/Fe]}=0.23\pm0.10$ dex, $\textmd{[Si/Fe]}=0.36\pm0.05$ dex, and $\textmd{[Ca/Fe]}=0.30\pm0.07$ dex, respectively. For comparison, \citet{fullton96} found values of $\textmd{[Si/Fe]}=0.68\pm0.13$ and $\textmd{[Ca/Fe]}=0.33\pm0.13$ dex. On the other hand, these elements seem to be overabundant from the estimates of \citet{gratton2015}; $\textmd{[O/Fe]}=0.53\pm0.09$ dex, $\textmd{[Mg/Fe]}=0.51\pm0.06$ dex, $\textmd{[Si/Fe]}=0.60\pm0.08$ dex, and $\textmd{[Ca/Fe]}=0.81\pm0.13$ dex. In this work, these authors speculate whether these high enhancements arise as an effect of age (NGC~6723 is very old, 12-13 Gyr, according to several age estimations; \citeauthor{marinfranch2009} \citeyear{marinfranch2009}, \citeauthor{dotter2010} \citeyear{dotter2010}, \citeauthor{vandenberg2013} \citeyear{vandenberg2013}), but they conclude that at least part of this trend can be due to a systematic effect that produces values from their HB sample that are higher by 0.2 dex  than those obtained from RGB stars \citep[as already pointed out in][]{carretta2010}.

The iron-peak element titanium presents a similar enhancement with $\textmd{[Ti/Fe]}=0.24\pm0.09$ dex. The same value is reported in \citet{fullton96}. Aluminium is also enhanced by $\textmd{[Al/Fe]}=0.31\pm0.21$ dex, while the s-process barium  by $\textmd{[Ba/Fe]}=0.22\pm0.12$ dex.

Metal-intermediate bulge globular clusters are old objects encoding important information about the environment conditions in which they were formed.  A detailed analysis of their chemical enrichment patterns can provide valuable information to understand whether or not they share a common origin. Furthermore, studies of other Galactic stellar components (bulge/disk) can benefit from the comparison with clusters as clean independent points in the age-metallicity-distance distributions.  

\begin{figure}
\begin{center}
\includegraphics[width=8.8cm]{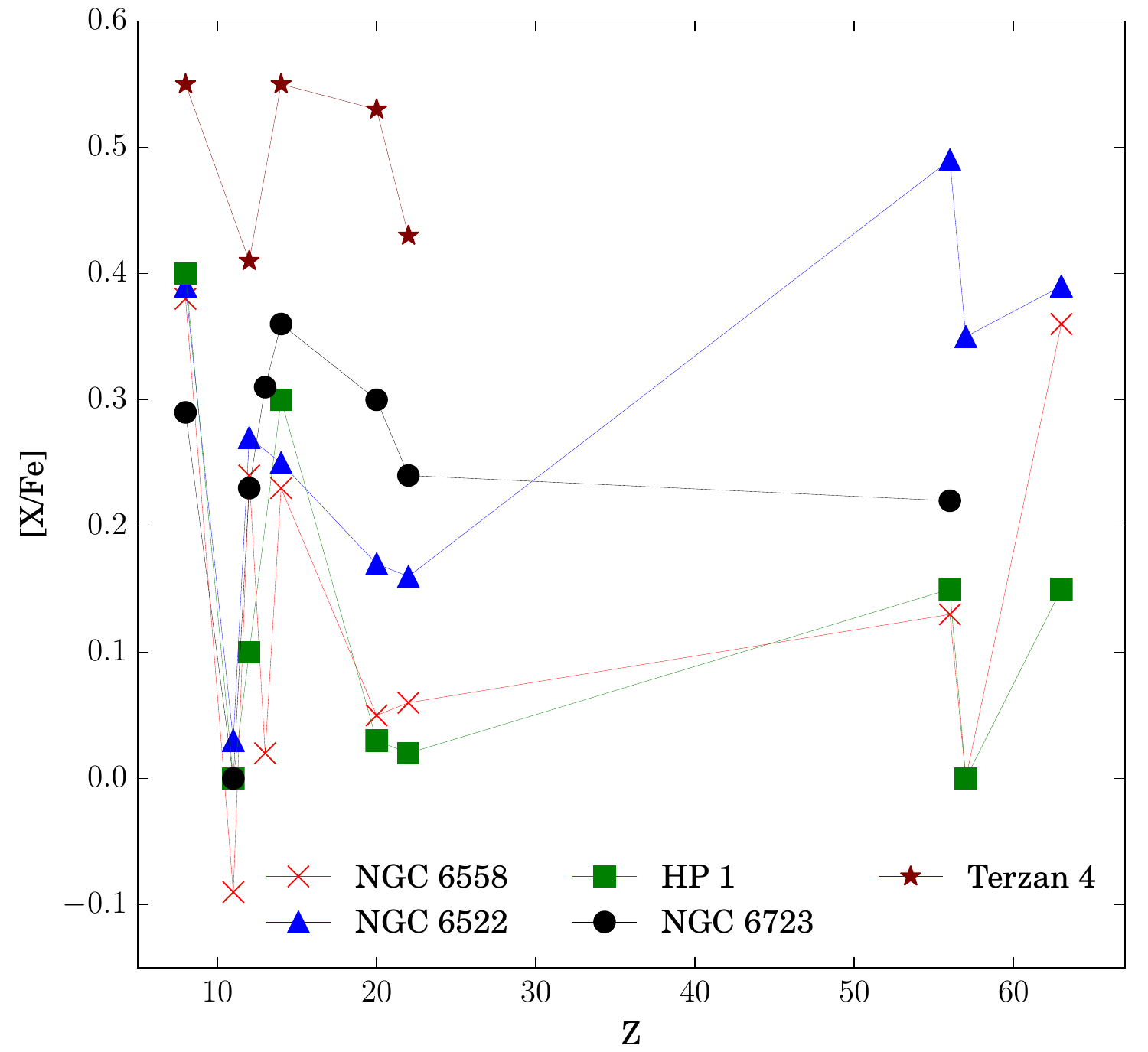}
\caption{Chemical enrichment pattern for NGC 6723 (black solid circles). For comparison, other metal-intermediate bulge globular clusters are represented: NGC 6558 (red crosses), NGC 6522 (blue solid triangles), HP1 (green solid squares), and Terzan 4 (black solid stars).} 
\label{fig:abund_pattern}
\end{center}
\end{figure}

In Fig. \ref{fig:abund_pattern}, the abundance pattern of NGC 6723 is compared with those of other intermediate-metallicity bulge globular clusters. A general similitude is present among NGC 6558 ($\textmd{[Fe/H]}=-0.97$ dex, \citeauthor{barbuy2007} \citeyear{barbuy2007}), HP 1 ($\textmd{[Fe/H]}=-1.0$ dex, \citeauthor{barbuy2006} \citeyear{barbuy2006}), and NGC 6522 ($\textmd{[Fe/H]}=-1.0$ dex, \citeauthor{barbuy2009} \citeyear{barbuy2009}). The abundance ratios of NGC 6723 are intermediate between those of the enhanced NGC 6522 and those of NGC 6558 and HP1. Such similarities, suggest a possible common origin for these clusters. In contrast, a more enhanced pattern is displayed by the metal-poor cluster Terzan 4. Given its low metallicity ($\textmd{[Fe/H]}=-1.60$ dex; \citeauthor{origlia2004} \citeyear{origlia2004}), and its somewhat higher velocity components with respect to the average of a sample of bulge globular clusters \citep{rossi2015}, it is possible that Terzan 4 is a halo cluster currently in the bulge region because it is close to its orbit pericenter. The $\alpha$-enhancement of Terzan 4 is high even compared to the field halo stars of \citet{nissen2010} and \citet{mikolaitis2014}, at the same metallicity.

\begin{figure*}
\begin{center}
\includegraphics[width=17cm]{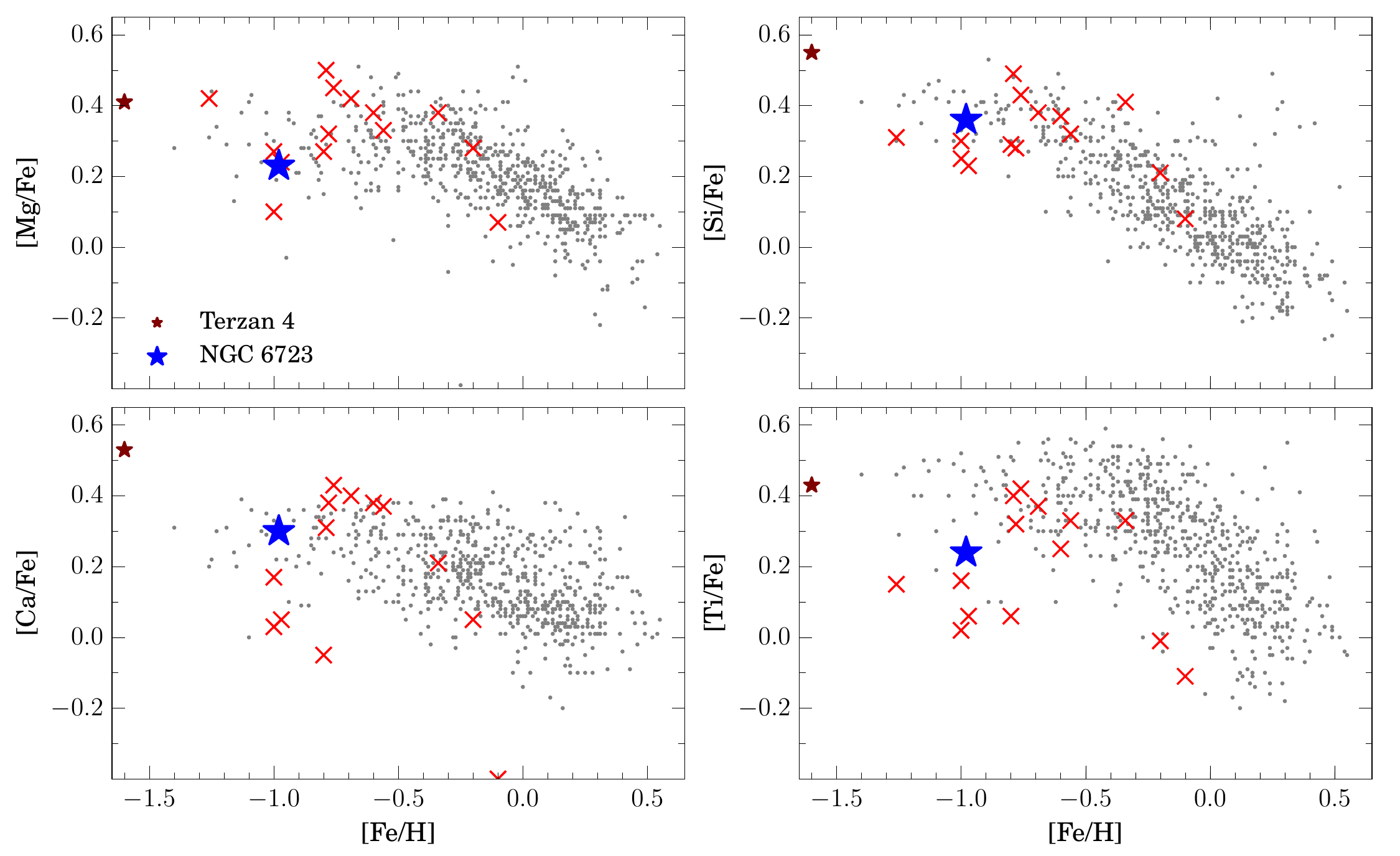}
\caption{[X/Fe] vs. [Fe/H]  for NGC 6723 compared with results from literature. Alpha-elements Mg, Si, Ca, and iron-peak element Ti are shown in the four panels of the figure. Gray dots depict the K giant bulge field stars from \citet{gonzalez2011}. A brown star stand for the metal-poor globular cluster Terzan 4 \citep{origlia2004}. Red crosses stand for a number of  metal-intermediate and metal-rich clusters from recent literature: NGC 6522 \citep{barbuy2009}, HP1 \citep{barbuy2006}, NGC 6558 \citep{barbuy2007}, NGC 6553 \citep{alves-brito2006}, NGC 6528 \citep{zoccali2004}, Terzan 1 \citep{valenti2015}, NGC 6441 \citep{gratton2007}, NGC 6388 \citep{wallerstein2007}, NGC 6624 and NGC 6569 \citep{valenti2011}, UKS1 and NGC 6539 \citep{origlia2005}, NGC 6440 \citep{origlia2008}, and NGC 6342 \citep{origlia2005}.} 
\label{fig:elem_ratio_context}
\end{center}
\end{figure*}

In order to put our derived abundances in the context of bulge studies, in Fig. \ref{fig:elem_ratio_context} we compare these abundances  with those of the K giant bulge sample presented in \citet{gonzalez2011}. As an extra comparison, several metal-rich and metal-intermediate globular clusters are also included from recent literature. The four panels show the [X/Fe] vs. [Fe/H] distributions for the alpha elements Mg, Si, and Ca, and Ti. In all cases, a general increase of [X/Fe] with decreasing metallicity is observed in the bulge field stars. 

The $\alpha$-element ratios of NGC 6723 are in good agreement with those of the bulge field stars. On the other hand, Ti is not following the $\alpha$-elements, and seems to be slightly less enhanced than in the bulge. This discrepancy is still compatible with the larger dispersion of [Ti/Fe] values at lower metallicity of the bulge sample.

The other clusters in Fig. \ref{fig:elem_ratio_context} compare well with the bulge sample. The agreement in Mg and Si is good in the entire metallicity range  of the sample. In the case of Ca and Ti, a fraction of the clusters seem to produce a sequence that is less enhanced and parallel to the bulge sequence. It would be of great interest to further explore this effect to figure out whether it is a spurious effect due to systematic errors, or it is the signature of a different, low alpha population. We emphasize that the comparisons discussed here are only valid  at a qualitative level. Indeed, literature determinations for clusters and bulge stars come from a heterogeneous set of studies with different analysis strategies. In this context, NGC 6723 seems to be characterized by an early prompt chemical enrichment, similar to that attibutted to the the metal-poor portion of the bulge.
Finally,  the metallicity we found for NGC~6723, of [Fe/H]$=-0.98$ dex, is comparable with the average values [Fe/H]$=-1.25$ dex \citep{kunder2008} and [Fe/H]$=-1.02$ dex \citep{pietrukowicz2012} determined for bulge RR Lyrae stars, as known tracers of old populations.

Further high resolution studies of individual stars in other bulge globular clusters might provide important clues to understand the place that they take in the general assembling of the Milky Way.  This might be especially important in the context of the current and future large spectroscopic surveys, exploring Galactic stellar population in increasingly large volumes. Such an observational effort can shade light on important points concerning formation mechanisms of stellar structures and chemodynamical evolution by providing constraints for Galaxy formation model predictions.

\begin{acknowledgements}
This work was partially supported by the BASAL CATA through grant PFB-06 CONICYT's PCI program through grant DPI20140066, the Chilean Ministry of Economy through ICM grant to the Millenium Institute of Astrophysics, and the Proyecto Fondecyt Regular 1150345.
\end{acknowledgements}


\bibliographystyle{aa}
\bibliography{biblio} 

\end{document}